\documentclass[showkeys, twocolumn, superscriptaddress]{revtex4-1}

\usepackage{amsmath}
\usepackage{amssymb}
\usepackage{bm}

\usepackage{fancyhdr}
\usepackage{graphicx}

\usepackage{color}
\usepackage{comment}

\definecolor{darkgreen}{rgb}{0,0.7,0.2}

\newcommand{\bq}{\begin{equation}}
\newcommand{\eq}{\end{equation}}
\newcommand{\ba}{\begin{eqnarray}}
\newcommand{\ea}{\end{eqnarray}}
\newcommand{\bc}{}

\newcommand{\const}{{\rm const}}

\graphicspath{{Figures/}}

\begin{document}

\author{Matthew A. A. Grant}
\affiliation{Cavendish Laboratory, University of Cambridge, JJ Thomson Avenue, CB3 0HE Cambridge, UK.}

\author{Bart{\l}omiej Wac{\l}aw}
\affiliation{SUPA, School of Physics and Astronomy, University of Edinburgh, James Clerk Maxwell Building, King's Buildings, Mayfield Road, Edinburgh EH9 3JZ, UK.}

\author{Rosalind J. Allen}
\affiliation{SUPA, School of Physics and Astronomy, University of Edinburgh, James Clerk Maxwell Building, King's Buildings, Mayfield Road, Edinburgh EH9 3JZ, UK.}

\author{Pietro Cicuta}
\affiliation{Cavendish Laboratory, University of Cambridge, JJ Thomson Avenue, CB3 0HE Cambridge, UK.}

\title{The role of mechanical forces in the planar-to-bulk transition in growing {\em{Escherichia coli}} microcolonies}


\begin{abstract}
Mechanical forces are obviously  important  in the assembly of three-dimensional multicellular structures, but their detailed role is often unclear. We have used growing microcolonies of the bacterium  \emph{Escherichia  coli}   to investigate the role of mechanical forces in the transition from two-dimensional growth (on the interface between a hard surface and a soft agarose pad) to three-dimensional growth (invasion of the agarose). We measure the position within the colony where the invasion transition happens, the cell density within the colony, and the colony size at the transition  as functions of the concentration of  the agarose. We use a phenomenological theory, combined with individual-based computer simulations, to show how mechanical forces acting between the bacterial cells, and between the bacteria and the surrounding matrix, lead to the complex phenomena observed in our experiments - in particular a non-trivial dependence of the colony size at the transition on the agarose concentration. Matching these approaches leads to a prediction for how the friction coefficient between the bacteria and the agarose should vary with agarose concentration. Our experimental conditions mimic
numerous clinical and environmental scenarios in which bacteria invade soft matrices, as well as shedding more general light on the transition between two- and three-dimensional growth in multicellular assemblies.
\end{abstract}

\maketitle

\section{Introduction}


The assembly of three-dimensional multicellular structures is a central theme in biology, from embryology to cancer \cite{wolpert1998principles,liotta_microenvironment_2001}. In general, this process is believed to be controlled by an interplay between mechanical forces between cells and their surrounding matrix~\cite{mammoto_mechanical_2010,marinari_live-cell_2012}, regulated cell growth (death and differentiation)~\cite{coucouvanis_signals_1995, gordon_differentiation_1994,asally_localized_2012}, biochemical interactions between cells~\cite{pera_integration_2003} and with the  matrix~\cite{hodor_cell-substrate_2000}, and cell migration~\cite{ridley_cell_2003}. However, dissecting the individual roles played by each of these factors remains largely an open challenge. Even from a purely mechanical point of view, our understanding of the forces exerted between cells, and between cells and their environment, remains limited.

An important stage in many multicellular developmental processes is the transition from two-dimensional to three-dimensional (2d to 3d) growth~\cite{wolpert1998principles}. Examples include eye formation in vertebrates, which begins as a ``bulge'' protruding from the surface of the ventrolateral forebrain~\cite{harris_pax-6:_1997}, and invasion of neighbouring tissues by skin cancer cells, initially confined to 2d by the basal membrane~\cite{liotta_microenvironment_2001}. For bacteria, an important transition from 2d to 3d growth occurs in biofilms which form on  surfaces such as water pipes~\cite{molin03},  or teeth, as well as in softer environments such as the skin~\cite{grice2009} or in foods~\cite{carpentier93}, and which are implicated in a variety of human pathologies~\cite{Singh03}. In the laboratory, bacterial microcolonies are often grown  trapped between a soft agarose pad and a glass slide, in order to  investigate gene regulatory and physiological processes at the single-cell level~\cite{Stewart2005}. In these experiments, cells initially grow in 2d but eventually invade the agarose to form multiple layers, hindering the tracking of individual cells.  It is likely that a similar transition occurs in bacterial biofilms growing at the interface between hard and soft materials such as surgical implants or catheters surrounded by tissue~\cite{donlan2002}.

\begin{figure*}[t!]
	\includegraphics[width=19cm]{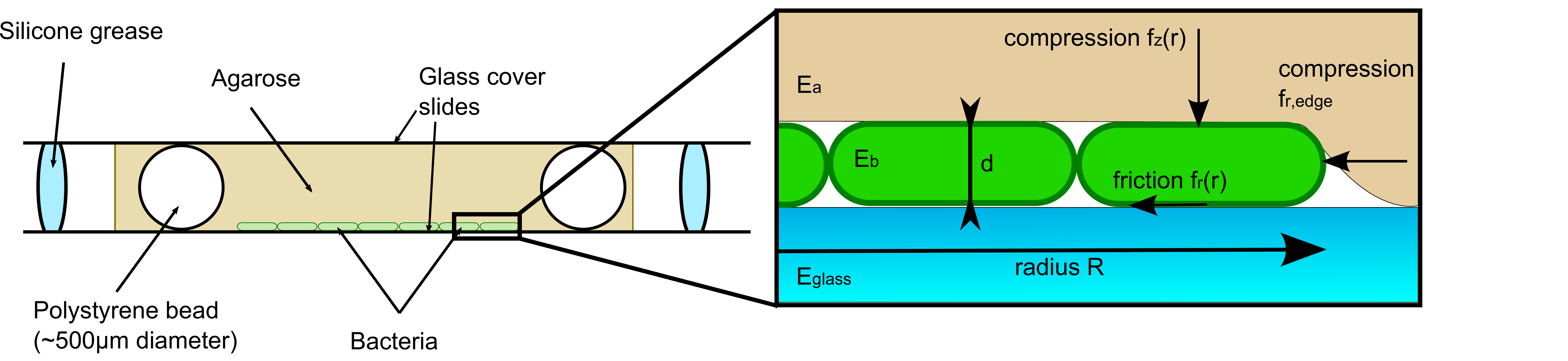}
	\caption{\label{fig:setup}{\bf Experimental setup.} A thin slab of LB-agarose is confined between two microscope slides. The agarose contains a small number of polystyrene beads which act as spacers and ensure that the thickness of the agarose slab is constant at $500\mu$m. The bacteria are pipetted onto the  top of the agarose before it is covered with the upper glass slide, and silicone grease is applied around the agarose to seal the sample. Zoomed-in region: illustration of the forces acting on the colony in our simulations. The microcolony is modelled as a flat disk of fixed height $d$ and variable radius $R$, which is compressed vertically (force per unit colony area $f_{z}(r)$) and radially (force per unit boudary area $f_{\rm r,\,edge}$) by the agarose, and radially by friction between the cells and the agarose (force per unit colony area $f_{r}(r)$).}
\end{figure*}

In this work, we investigate the role of mechanical forces in the transition from 2d to 3d growth for {\em{Escherichia coli}} bacteria sandwiched between a glass slide and an agarose gel (Figure~\ref{fig:setup}),  using experiments, phenomenological theory, and computer simulations. Under these conditions, bacteria proliferate to form microcolonies, which are initially confined to the surface of the agarose but eventually invade the agarose to form a 3d community. In these microcolonies, the cells do not demonstrate any  active motility (such as swimming, swarming, or gliding~\cite{Henrichsen1972,Harshey2003}), but they do move due to ``pushing" interactions with other cells as they proliferate. This system provides a useful  simple model for 3d multicellular assembly, since relatively few factors are at play. We show that the elasticity of the substrate and of the bacteria (which are compressed and bent by the interaction with the substrate) plays an important and non-trivial role in determining the colony size at which this transition happens. Our results can be explained purely by mechanical forces, with some simple assumptions about the nature of the friction between the bacteria and the agarose, and lead to a prediction for the dependence of this friction coefficient on the agarose concentration. More generally, our work should lead to better understanding of the invasion of soft materials by bacteria, and shed light on how mechanical interactions between cells and their environment can lead to the emergence of complex 3d structures.

\begin{figure}[t!]
	\includegraphics[width=8.2cm]{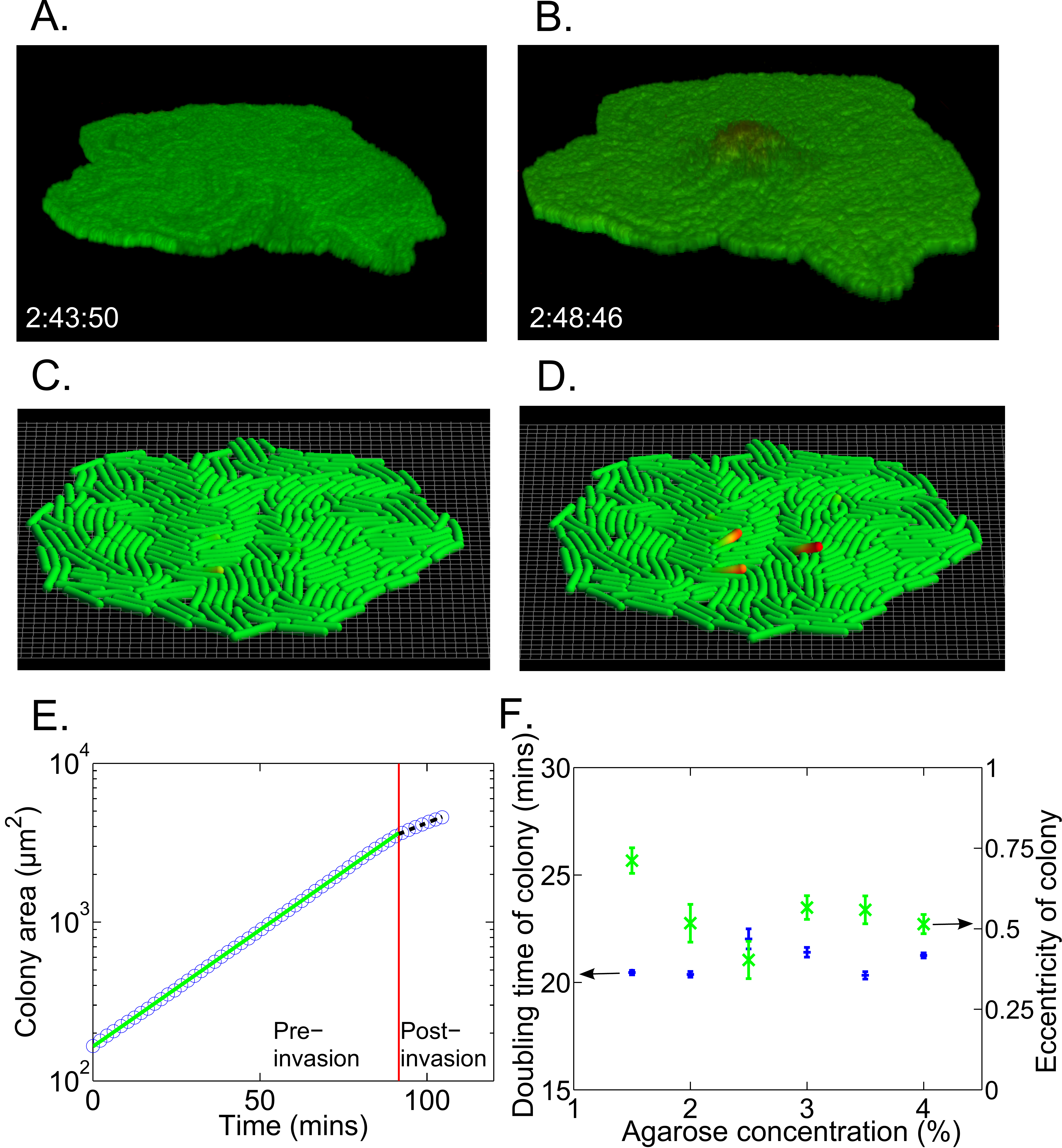}
	\caption{\label{fig:transition}\textbf{The transition from 2d to 3d growth in {\em E. coli} microcolonies.}  Confocal images of a microcolony just before (A) and after (B) the invasion; the microcolony is shaded for depth, showing in dark where invasion has occurred.  Simulations of a microcolony show the same phenomenon: the in-silico colony is represented before (C) and after (D) invasion;  bacteria are shown in dark if they have invaded the agarose. An example of microcolony area as a function of time is shown in (E): The area first increases exponentially (straight segment of the plot), but upon invasion the observed area growth rate decreases. The solid line is the best fit to the initial exponential growth phase, combined to a `post-invasion' phase. The vertical line is at the fitted invasion time, as determined by the best fit of the solid and dashed lines. The fit has been performed on the two segments with a moving crossover. F: The doubling time of the microcolony, obtained from the preinvasion area growth rate, is unaffected by agarose concentration.}
\end{figure}

\section{Materials and Methods}

\subsection{Experimental methods}

{\it Agarose preparation.}
 Agarose (Sigma Aldrich - A9539) was mixed with Luria Bertani (LB) broth powder (Sigma Aldrich - L3022, containing 10 g/l tryptone, 5 g/l yeast extract and 5 g/l NaCl) and water (to make a final LB powder concentration of 2\% wt. This concentration of LB broth powder is optimal as it enables \textit{E. coli} to double at their fastest growth rates) and then autoclaved. A range of different concentrations of agarose was explored. Once the autoclaving cycle was complete, the agarose-LB mix was stirred for one hour at 80$^\circ$C, to ensure thorough mixing, and then stored at a temperature ($  \approx 50^\circ$C) that kept the solution in a liquid form. The range of agarose concentration studied was from $  1.5\%$wt to $  4\%$wt: This is the widest possible range in these experiments, since below $  1.5\%$ the agarose becomes too soft and the bacteria become motile, while above  $  4\%$wt the very high viscosity prevents  sample preparation.

{\it Bacterial strain and growth conditions.}
The fluorescently tagged \textit{Escherichia coli} K-12 strain BW25113+PKK\_PdnaA-GFP (which contains the gene for GFP under the control of the {\em dnaA} promoter, on a plasmid) was grown overnight in 5\,ml~LB supplemented with 5\,$\,\mu$l of 100 mg/ml ampicillin (to make a final concentration 100\,$\,\mu$g/ml) at 37$^\circ$C. The overnight culture was diluted into fresh LB at a ratio 1:300, one hour before use;   the bacteria were in exponential phase when the experiment commenced.

{\it Sample preparation.}
 Trace amounts of polystyrene beads of diameter 500$\,\mu$m  were mixed with the liquid agarose in an Eppendorf tube. 1\,ml of this agarose-bead mixture was then pipetted onto a cover slide, compressed with a second cover slide and allowed to dry.  A small section of this agarose was then cut and placed on a new microscope cover slide. 2\,$\,\mu$l of the bacterial culture (in exponential phase) was then pipetted onto the agarose in a single drop. The slide around the agarose was covered in silicone grease (to seal the sample), and a cover slide was placed on top. The sparse 500$\,\mu$m beads act as spacers and ensure that the agarose layer  between the coverslip and microscope slide remains of constant depth during the imaging process.

{\it Data collection.}
The sample was imaged using a Leica SP5 confocal microscope, with an automated stage allowing for several microcolonies to be tracked over time. To obtain the area of the microcolony (which was measured at intervals of one minute), the microscope's transmitted beam was recorded (equivalent to  bright field imaging). In some experiments, to obtain 3d images of the microcolony, the microscope was used in confocal mode in order to build up a z-stack.
All experiments were performed at 37$^\circ$C.

{\it Data analysis.}
Each bright field image was analysed using custom scripts written in Matlab:  a simple intensity threshold on the image was sufficient to distinguish the colony from the background, and hence the area and shape could be calculated; watershed filtering was used to segment individual cells in selected experiments.

{\it Detection of the invasion transition.}
The time at which the microcolony invaded the agarose was identified by finding the discontinuity in its area growth rate (see Figure~\ref{fig:transition}E), which was found (from our confocal images) to correspond to the moment at which the first cell escapes from the 2d layer. The invasion transition can also be  identified by eye in our brightfield images, but in these images a small shift in focus can prevent  the first invading cell from being spotted immediately. Use of confocal imaging in all our experiments, while accurate, would have been much more  resource intensive.

\subsection{Simulations}

{\it Individual-based model for elastic bacteria.}
Our computer simulations were implemented in a purpose-written C++ program. Individual bacteria were modelled as elastic rods, represented by $n_{\rm spheres}=8-16$ overlapping spheres (with new spheres added as the rod grows), linked together via nonlinear springs, such that the rod dynamics was described by Euler-Bernoulli dynamic beam theory \cite{han_dynamics_1999}. Repulsive interactions between bacterial cells were modelled  according to the Hertzian theory of contact mechanics \cite{landauelasticity}, by imposing a repulsive potential between spheres located in different bacterial rods; equivalent forces were also used to represent the interaction between the cells and the solid glass surface. In addition to these repulsive forces, cells also experienced  dry static frictional forces (according to Amonton's laws of friction, with  friction coefficient $0.3$); these were assumed to act between any pair of cells that were in contact. The frictional forces acted in the direction opposite to the local sliding velocity. In some simulations we also used Stokes-like, velocity-dependent friction, as described in Appendix E.

{\it Bacterial growth.}
Cell growth was modelled by a linear expansion of the rod length; upon reaching twice its initial length, each rod was split in two, producing two equal-sized daughter cells. We assumed the average length of an uncompressed bacterium to be $5\mu$m, in agreement with our experimental data for low agarose concentrations. We included stochasticity in the growth dynamics by allowing the growth rate of individual cells to vary by  $  \pm 10$\% around a mean doubling time of 20\,min  (this meant that the simulation lost synchrony in cell division  after about 10 generations, as has been observed experimentally~\cite{hoffman_synchrony_1965}). Nutrients were not modelled explicitly (the mean growth rate was assumed to be the same for all cells).

{\it Interactions between bacteria and agarose.}
The agarose was modelled implicitly in our simulations, via vertical and horizontal compression forces, and horizontal frictional forces, acting on the bacteria. In our simulations of single microcolonies, vertical compression of the bacteria by the agarose was represented by a force of magnitude $N=(a_{\rm cell}/n_{\rm spheres})E_{\rm a}d/(\pi \sqrt{R^2-r^2})$, acting on each of the $n_{\rm spheres}$ spheres making up a bacterial rod (cf. Eq.~(\ref{eq:vert})). Here, $a_{\rm cell}$ is the horizontal area of a given bacterium (which can change due to growth or compression), $r$ is its radial distance from the microcolony centre,  $R$ is the radius of the microcolony, $d=1.4\mu$m is the height of the microcolony (this is a typical cell diameter as measured in our experiments) and $E_{\rm a}$ is the Young elastic modulus of the agarose. Radial compression by the agarose was modelled by an inward radial force $(a_{\rm cell}/n_{\rm spheres})E_{\rm a}d/(2\pi R)$, acting on spheres that form part of bacteria located at the periphery of the microcolony. The theoretical arguments leading to these force functions are given in the Results section and in Appendix A. For simulations of colliding microcolonies, different force functions were used (for details see Appendix C).

In all our simulations, friction between the bacteria and the agarose, and between the bacteria and the glass surface, was represented by a force $  F=k N$ acting on each sphere in the direction opposite to its velocity, where $  k$ is the friction coefficient, and $N$ is the vertical compression force acting on that bacterium.
We explored different functional forms for the dependence of $k$ on the concentration $C_{\rm a}$ of the agarose, as discussed in the Results section. In our simulations, we assumed that the frictional force does not depend on the velocity $v$ of cells relative to the surface. This is different to previous works~\cite{volfson_biomechanical_2008,boyer_buckling_2011,haseloff2012} which have assumed  Stokesian-like friction, proportional to the cells' velocity. In our opinion, this static friction model is more physically realistic; however, we show in Appendix E that repeating our simulations with dynamic Stokesian friction $F=k v N$ produces almost identical results.

{\it Elasticity parameters.}
The elastic modulus $E_{\rm b}$ of an {\em{E. coli}} cell in our experiments is not known (existing data varies between $0.1-200$\,MPa \cite{agar_E}); in our simulations we found that  $E_{\rm b}=375$\,kPa gave a good fit to our experimental data. The  elastic modulus of glass was set to $E_{\rm glass}=10$\,MPa (i.e. only 25 times larger than that of the bacteria): this  represents a compromise between speed and realism of the simulation \footnote{In reality, $  E_{\rm glass}$ is $  10^3 - 10^5$ larger than $  E_{\rm b}$, but this would require impractically small time steps in the numerical algorithm}, and is large enough to prevent cells from invading the glass slide. We varied the elastic modulus $E_{\rm a}$ of the agarose in the range $100$\,kPa to $800$\,kPa, in agreement with the experimentally explored range of agarose concentrations.

{\it Triggering the 2d to 3d transition.} In the growing colonies, we expect that the invasion transition happens when the ``squeezing" of the microcolony due to radial (friction and/or compression) forces causes cells to move out of the horizontal plane, overcoming the vertical forces resulting from compression of the agarose. The translation of horizontal radial forces into vertical motion  happens because of  small local inhomogeneities in the surface and/or in the shape of individual bacteria, as well as the Euler buckling instability in compressed rod-shaped bacteria. To reproduce this in our simulations, we introduced local inhomogeneities in the glass/agarose surfaces, represented by a sinusoidal modulation of the glass' height with 10\,nm amplitude and period 1\,$\mu$m along both horizontal axes ($x$ and $y$).

{\it Dynamics.} The dynamics of the system was modelled by solving Newton's equation of motion for the spheres, with the only source of damping coming from the frictional forces. We used a simple Euler method to integrate the system of differential equations for the position and velocity of each sphere with a fixed time step $  dt = 2^{-18} - 2^{-16}$\,h.

{\it Detecting the 2d to 3d transition.}
In our simulations, the 2d to 3d transition happens very rapidly and can  be detected accurately by measuring the vertical distance of the surface of each of the spheres in each bacterium from the glass surface (defined as the smallest distance between the surface of the sphere and  the glass surface). The transition was defined to be the moment at which this distance was more than $ (1/2)d = 0.7\,\mu$m for any sphere.

\section{Results}

\subsection{Experimental observations}

\subsubsection{Bacterial microcolonies undergo a sharp ``invasion'' transition from 2d to 3d growth}

Tracking the microcolony area as a function of time in our experiments reveals that the colony grows exponentially throughout our experiments (suggesting that there is no nutrient limitation), but with a discontinuity in the area growth rate (Figure~\ref{fig:transition}E) which occurs after  80-90 minutes of growth for an agarose concentration of 3\%. Confocal microscopy images (see Fig.~\ref{fig:transition}A, B) show that this  discontinuity corresponds to the formation of a second layer of bacterial cells, i.e. the transition from 2d to 3d growth.

\subsubsection{Invasion of the agarose first occurs near the  microcolony centre}

\begin{figure*}[t!]
	\includegraphics[width=16cm]{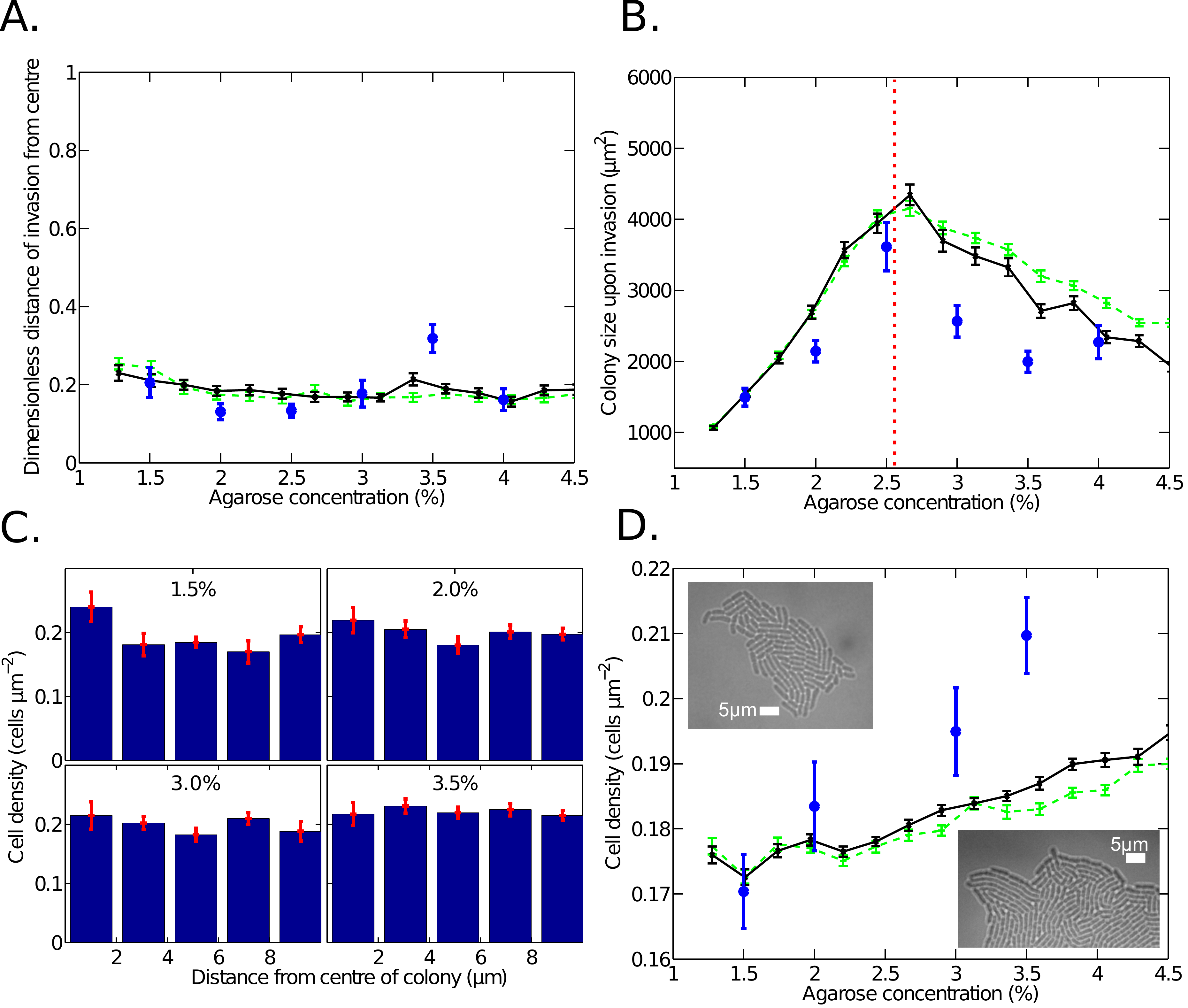}
	\caption{\label{fig:results}\textbf{Simulations using only mechanical forces are able to match experimental observations.} A: The dimensionless distance of invasion from the centre of the colony. Markers show experimental data; lines show simulation results where friction is given by Eq.~(\ref{eq:hor}) for $\alpha=0.4$ (light, dashed) and $\alpha=1$ (dark, solid), with the friction coefficient from Eq.~(\ref{eq:kk}). B: The area of the colony at which invasion first takes place. Data markers are the same as in panel A. The vertical line marks the agarose concentration $C_{\rm a}=2.55$\% corresponding to $E_{\rm b}=375$\,kPa, which is very close to $364$\,kPa ($2.5\%$wt) at which the friction coefficient is assumed to stop decreasing.  C: The density of the bacteria in the colony shows no dependence on the radial position. D: The mean density of bacteria in the colony increases with agarose concentration. Images correspond to colonies at agarose concentrations of 1.5\% (left) and 3.5\% (right). Number of colonies analysed for panels A and B is between 9 and 24 for each agarose concentration, while for panels C and D between 6 and 15 colonies are analysed for each agarose concentration shown. There are typically 250 bacteria in the colonies at 1.5\% when invasion occurs (approximately 8 generations) whereas for 3\% there are typically 500 bacteria (approximately 9 generations). \\
}
\end{figure*}

The colonies are roughly circular-symmetric. We observe that invasion of the agarose by the bacteria consistently starts very close to the centre of the microcolony. Using confocal microscopy, we were able to pinpoint the exact position $\mathbf{x}_{inv}$ within the microcolony where the first bacterium escapes from the 2d microcolony to form a second vertical layer.   In order to compare microcolonies of different sizes, we define the dimensionless distance
\bq
	d_{inv} = \frac{|\mathbf{x}_{inv} - \mathbf{x}_C|}{\sqrt{A/\pi}}, \label{eq:d_B}
\eq
where  $\mathbf{x}_C$ is the centre of mass of the microcolony and  $A$ is its area. Thus $d_{inv}=0$ if invasion happens at the centre of the microcolony and $d_{inv}=1$ if it happens at the very edge of the microcolony. Our results show that invasion always occurs near the centre of the microcolony, with $d_{inv}\approx 0.2$ (Figure~\ref{fig:results}A), over a wide range of agarose concentrations.

\subsubsection{Invasion does not require secreted factors}

The fact that invasion occurs near the colony centre might suggest that it is triggered by biochemical factors secreted by the cells, which might affect their mechanical behaviour  (for example secreted factors can affect motility in \textit{Legionella pneumophila}~\cite{Stewart2009} and \textit{Pseudomonas aeruginosa}~\cite{Fauvart2012}), or the physical structure of the agarose~\cite{tipler_1985}. Such factors would be expected to be present at the  highest concentration, and for the longest time, near the microcolony centre.
To test this hypothesis, we tracked the invasion transition in {\em colliding microcolonies} - i.e. microcolonies that originate from closely spaced cells and collide as they expand in 2d. In these experiments, we would still expect accumulation of secreted factors near the centres of the original individual microcolonies. Importantly, however, we observed that  invasion of the agarose often occurred at the contact point between colliding colonies, rather than at their original centres (Fig.~\ref{fig:collision}A).  This suggests that secreted biochemical factors are unlikely to be playing an important role in the invasion transition.

\subsubsection{The invasion transition shows a complex dependence on agarose concentration}

To probe the mechanical forces acting within the microcolonies and their role in the invasion transition, we manipulated the elasticity of the agarose by varying its  concentration. The elastic modulus of the agarose has been related to its concentration (see the supplementary material of Ref.~\cite{agar_E}) by
\bq
	E_{\rm a} \approx 216C_{\rm a}-176,	\label{eq:agar_e}
\eq
where $C_{\rm a}$ is the agarose concentration in \%wt and $E_{\rm a}$ is the elastic modulus in kPa. Since our experiments were conducted for $C_{\rm a}=1.5\% - 4\%$, we expect that the elastic modulus $E_{\rm a}$ ranges from $\approx 150$\,kPa to $\approx 700$\,kPa \footnote{The exact relationship between $E_{\rm a}$ and $C_{\rm a}$ can depend on the agarose type, preparation method etc. The agarose used in our experiment was very similar to that of Ref.~\cite{agar_E} and hence we expect that Eq.~(\ref{eq:agar_e}) holds to a good approximation.}.

Focusing first on the basic properties of the microcolonies prior to the invasion transition, we find that increasing the agarose concentration has no effect on the pre-invasion growth rate (Figure~\ref{fig:transition}F) but does increase the average density of cells within the growing colony (Figure \ref{fig:results}D). For all agarose concentrations, the cell density is rather uniform throughout the colonies  (Figure \ref{fig:results}C).

Interestingly, the microcolony size at which the invasion transition happens shows a complex dependence on the agarose elasticity: For low agarose concentrations, the colony  area at invasion  increases approximately linearly with $C_{\rm a}$, but for higher agarose concentrations $C_{\rm a}$ the area at invasion actually {\em{decreases}} as $C_{\rm a}$ increases (Figure~\ref{fig:results}B). The peak in the area at invasion occurs at an agarose concentration of $C_{\rm a}=2.5\%$wt, corresponding to an elastic modulus $E_{\rm a} \approx 360$\,kPa.

\subsection{Mechanical theory}

Our experimental observations can be understood by considering the nature of the forces acting on the bacteria within the microcolony. We assume that these forces consist of (i)~repulsive forces between the bacteria (ii)~frictional forces between neighbouring bacteria, and between bacteria and the glass and agarose surfaces and (iii)~elastic forces exerted by the agarose on the bacteria due to its compression by the microcolony as it grows.  In section~\ref{section:individual-based} we will show that individual-based computer simulations including these force contributions can account for the phenomena observed in our experiments. First, however, we consider in detail the nature of the elastic and frictional forces arising from the interactions between the microcolony and the agarose, and their implications for how the microcolony area at the transition should scale with the agarose concentration.

\begin{figure}[t!]
	\includegraphics[width=8.5cm]{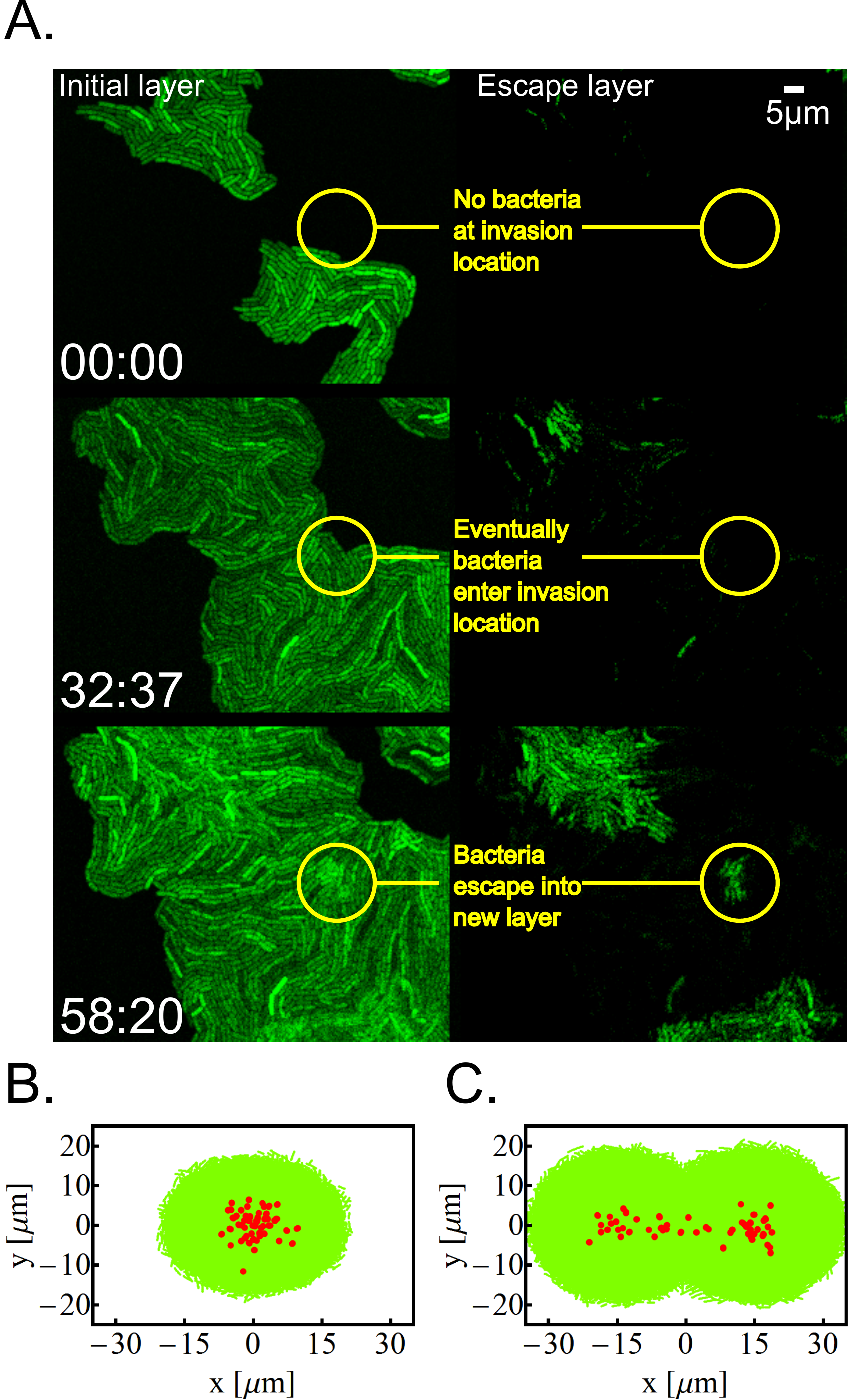}
	\caption{\label{fig:collision}\textbf{Invasion does not require  chemical factors.} A: Colonies that collide as they grow are shown at three different times (as labeled min:sec), and at two focal planes separated by 2.1\,$\mu$m. At the time of the first image, initially separated colonies have already collided, but bacterial escape has not yet occurred anywhere. As the collision proceeds, the second layer of cells begins to form at a point denoted by the red circle. Invasion thus occurs in the area previously not occupied by bacteria. This implies that invasion is not triggered by chemical cues because chemical factors secreted by the bacteria could not have accumulated at the collision site. The locations at which invasion first occurs in colliding colonies can be simulated:  10\,$\mu$m (B) and 30\,$\mu$m (C) separation are set between the two cells that initiate the colonies.  The panels show the position of the invasion transition (solid dots), and the bacteria (small light segments), overlaying 50 individual simulations. It can be seen that invasion often  occurs around the point $(0,0)$ where the two colonies collide (with a wide distribution). }
\end{figure}

\subsubsection{Forces due to compression of the agarose}

As illustrated in Fig.~\ref{fig:setup}, the growing microcolony  compresses the agarose layer, which will therefore exert elastic forces on the bacteria. To calculate these forces, we assume that the microcolony can be treated as a rigid disk of radius $R$ which is pressed into the agarose to a depth $d\approx 1.4\,\mu$m (this being a typical cell diameter in our experiments)\footnote{Although, as we show, the elastic modulus of the bacteria is comparable to that of agarose, Eq.~(\ref{eq:vert}) shows that for $R\gg d$ the compression force per unit area of microcolony $f_{\rm z}$ is much smaller than $E_{\rm b}$ for all but $r$ close to the colony boundary.}. From the theory of contact mechanics~\cite{landauelasticity}  the vertical compression force per unit area $f_{\rm z}(r)$ (see Fig.~\ref{fig:setup}), at a distance $r$ from the centre of the microcolony, is:
\bq
	f_{\rm z}(r) \approx \frac{E_{\rm a}d}{\pi \sqrt{R^2-r^2}}. \label{eq:vert}
\eq
There will also be a radial contribution to the compression force, acting on bacteria at the edge of the microcolony.  In Appendix A, we show that the magnitude of this force (per unit area) is
\bq
f_{\rm r,\,edge}\approx E_{\rm a}d/(2\pi R) \label{eq:rad}
\eq
however, as discussed below we do not expect this to play a major role. The quantities $f_{\rm z}$ and $f_{\rm r,\,edge}$ are in units of kPa and can be thought of as normal stresses specified at the boundary of the colony.

\subsubsection{Forces due to friction with the agarose}

We also expect frictional forces, acting in the inward radial direction, to exist between the growing microcolony and the agarose and glass surfaces (see Fig.~\ref{fig:setup}), with  magnitude proportional to the vertical compression stress $f_{\rm z}$. Previous observations of the surface friction of polymer gels~\cite{gong_surface_2002} suggest that the friction force per unit area $f_{\rm r}(r)$ can be generically expressed as
\bq
	f_{\rm r}(r) = k f_{\rm z}^{\alpha}(r) E_{\rm a}^{1-\alpha}, \label{eq:hor}
\eq
where $\alpha$ is an exponent which depends on the chemical structure of the gel ($0 < \alpha < 1$), and $k$ is a dimensionless friction coefficient which we expect to depend on the concentration of the agarose (note that in the case where $k$ is a constant and $\alpha=1$, equation (\ref{eq:hor}) reduces to Amonton's first law of friction $f_{\rm r}=k f_{\rm z}$, or $F=kN$~\cite{gao_frictional_2004}).

In our experiments, we believe that this radial friction force plays a more important role than the radial elastic force which arises directly from compression of the agarose. This is because, in our experiments (with isolated microcolonies), invasion of the agarose occurs very close to the microcolony centre  (Fig.~\ref{fig:results}A). This is very well reproduced by our computer simulations  in which the radial elastic force is much smaller than the frictional forces (see also the theoretical arguments in the next section). In contrast, when we make the frictional forces much smaller than the radial elastic force, we see that invasion occurs further from the centre (see Appendix~D and Figure~\ref{fig:rad_only} therein), which is incompatible with the experimental data.

\subsubsection{Expected consequences for the friction coefficient}

Building on the preceding theoretical arguments for the nature of the elastic and frictional forces, we now speculate briefly on the likely form of the friction coefficient $k$.

For simplicity, let us consider a one-dimensional analogue of a microcolony: a chain of $n$ bacteria sandwiched between an agarose surface and a rigid substrate. We assume that the first bacterium is fixed at $x=0$, so that as the bacteria proliferate, the chain extends along the positive semiaxis.
The bacteria are compressed vertically with force per unit area $f_{\rm z}(x) = E_{\rm a}d / (\pi \sqrt{L^2-x^2})$, in analogy with Eq.~(\ref{eq:vert}) (where $L$ is the length of the chain). The transition happens when  the chain of bacteria ``buckles'' in the vertical direction; we expect that this occurs because small inhomogeneities translate inward horizontal forces within the chain into vertical forces. Once these become large enough to overcome the vertical compression force, the bacteria invade the agarose.

As the bacteria proliferate, the chain expands in the positive $x$ direction, and the bacteria experience frictional forces acting in the opposite direction. These forces will be transmitted along the chain so that the maximal horizontal stress is experienced by the first bacterium in the chain (at $x=0$). We can calculate the total horizontal force on this bacterium as
\bq
	F_{\rm x,\,total} = d\sum_{i=1}^n \int_{x_{i-1}}^{x_i} f_{\rm x}(x) dx = d\int_0^L f_{\rm x}(x) dx, \label{eq:ftot}
\eq
where $f_{\rm x}(x)=k f_{\rm z}^{\alpha}(x) E_{\rm a}^{1-\alpha}$ is the frictional force per unit area (by analogy with Eq.~(\ref{eq:hor})), $\{x_i\}$ are the positions of the contact points between the bacteria, and $d$ is the width of a bacterium.

This  force is transformed into a vertical force pushing the bacteria into the agarose, by factors such as irregularities in the agarose (or glass) surface, differences in the diameters of individual bacteria, or Euler buckling of individual bacteria. We represent this by supposing a vertical force component $a F_{\rm x,\,total}$, where $a\ll 1$ is assumed to be a constant independent of $C_{\rm a}$.

The bacterium at $x=0$  will penetrate the agarose if  this vertical force component, directed into the agarose, is greater than the vertical compression force (which is directed away from the agarose).
This condition, together with Eq.~(\ref{eq:ftot}), leads to the following expression for the critical chain length $L$ at which the transition happens:
\bq
	L^{2-\alpha} = \frac{{\rm const}}{a k}. 
	\label{eq:balance}
\eq
In Eq.~(\ref{eq:balance}), the  explicit dependence on the elastic modulus of the agarose $E_{\rm a}$ has cancelled out. Thus, any dependence of the critical chain length $L$ on the agarose concentration must arise implicitly from a dependence of the friction coefficient on $C_{\rm a}$.

In our experiments, microcolonies are of course two-dimensional. While it is possible to construct a similar argument for the radial frictional stress field inside a circular microcolony, it is not clear whether the translation of horizontal into vertical stress can be represented in such a simple way, and to make progress we need to turn to computer simulations (see next section). Nevertheless, our arguments for the one-dimensional bacterial chain do suggest that the strong dependence of the critical microcolony area on the agarose concentration which we observe in our experiments is likely to originate from a concentration dependence of the friction coefficient.

\subsection{Computer simulations}
\label{section:individual-based}

To demonstrate that mechanical arguments can indeed explain the full range of phenomena that we see in our experiments, we carried out individual-based computer simulations, in which elastic, rod-shaped bacteria grow, divide, and interact mechanically with each other and with the agarose and glass surfaces. The agarose is represented implicitly by position-dependent vertical and horizontal forces acting on the bacteria, as predicted by our theoretical arguments (Eqs.~(\ref{eq:vert})-(\ref{eq:hor}); see Methods for details of the implementation).


Most parameters in our simulation have been either measured by us or taken from the literature (see Methods). We fix the two remaining parameters, the elastic modulus of the bacteria $E_{\rm b}$, and the value of the friction coefficient $k(C_{\rm a})$ at one specific agarose concentration, by comparing our simulations with the experimentally determined area and cell density at the transition. We choose $1.5$\% agarose as the reference point; this corresponds to $E_{\rm a}= 148$\,kPa \cite{agar_E}. Performing simulations for many different $E_{\rm b}$ and $k(C_{\rm a}=1.5\%)$, we find that the area and cell density match those determined experimentally for $1.5$\%wt agarose if we assume $E_{\rm b}\approx 375$kPa, and $k(C_{\rm a}=1.5\%)\equiv k_0=0.7$ (see Figs.~\ref{fig:results} and \ref{fig:sf1}). Interestingly, this automatically leads to the reduced distance from the centre at which the transition happens matching the experimental data quite well.

The 1d theory developed in the previous section suggests that assuming $k=\const$ should lead to a constant colony area upon transition, regardless of the agarose concentration. In fact, we find that our simulations (see Appendix B) show, that when we take $k=\const$, the area at the transition actually decreases slowly with $C_{\rm a}$ \footnote{This is probably caused by a more complicated distribution of the forces in the 2d colony as compared to the 1d case, and by the compression and bending of cells.}. This behaviour cannot match the results of our experiments, suggesting, as we deduced from the 1d theory, that $k(C_{\rm a})$ must depend on the agarose concentration.

It is therefore necessary to find a functional form of $k(C_{\rm a})$ that can match our experimental data for all $C_{\rm a}$. More specifically, we would like to find $k(C_{\rm a})$ such that the colony area at the transition first increases linearly with $E_{\rm a}$ (expressed as a function of $C_{\rm a}$ by Eq.~(\ref{eq:agar_e})), reaches a peak at about $2.5$\% agarose, and then decreases again for larger concentrations. In Appendix B we show that good agreement for concentrations up to $2.5$\% can be obtained if we assume $k(C_{\rm a}) = k_0 [148/E_{\rm a}(C_{\rm a})]$, where $E_{\rm a}(C_{\rm a})$ is expressed in kPa (via Eq.~(\ref{eq:agar_e})) and the proportionality factor $k_0=0.7$ has been chosen so that the area upon transition matches the experimentally determined area for $1.5$\% agarose.

To reproduce the peak and the decrease in the area at the transition for higher agarose concentrations, we assume that the friction coefficient $k(C_{\rm a})$ first  decreases with the agarose concentration as $\sim 1/E_{\rm a}(C_{\rm a})$ (see above), and then becomes constant above the concentration $2.5$\%, which corresponds to $E_{\rm a}=364$\,kPa:
\bq
	k(C_{\rm a}) = \left\{ \begin{array}{cc} k_0 \times 148{\rm \,kPa}/E_{\rm a}(C_{\rm a}) & {\rm for}\; C_{\rm a}\leq 2.5\% \\ k_0 \times 148{\rm \,kPa}/364{\rm \,kPa} & {\rm for}\; C_{\rm a}>2.5\% \end{array} \right. \label{eq:kk}
\eq
Interestingly, the value of $E_{\rm a}$ at which the coefficient stops decreasing is very close to the assumed elastic modulus of the bacteria $E_{\rm b}=375$\,kPa which was necessary to reproduce the experimental data for $1.5$\% agarose. This functional form of the friction coefficient provides good agreement between the simulated and experimentally determined area at the transition (Fig.~\ref{fig:results}B). Moreover, for the same set of parameters, our simulation results also agree remarkably well with the experimental data for the distance from the centre of the colony at which invasion happens (Fig.~\ref{fig:results}A), and reproduce the experimental trend in the density of the bacteria as a function of  $C_{\rm a}$ (Fig.~\ref{fig:results}D) - although here the quantitative agreement is slightly less good. Repeating these simulations for two different values of $\alpha$ (see Eq.~(\ref{eq:hor})); $\alpha=1$ (Amonton's law of friction) and $\alpha=0.4$ - the value found experimentally for agarose in contact with glass \cite{gong_surface_2002}, we find that both choices produce almost identical results (Fig.~\ref{fig:results}).

The formula (\ref{eq:kk}) should not be taken as a quantitative prediction; rather, it only indicates the trend in the friction coefficient that is necessary to match our experimental observations. While further experiments are needed to measure the friction coefficient of a bacterial cell on an agarose surface, such a non-monotonic dependence on the agarose concentration would not be entirely unexpected, since the friction coefficient between polymer gels and macroscopic surfaces has been observed to vary significantly with applied load (increasing or decreasing, depending on the gel type and its degree of swelling), and to depend in a non-trivial way on the polymer concentration~\cite{gong_surface_2002}.

We have also performed simulations in which two microcolonies grow close together in space, such that they collide prior to the invasion transition. In these simulations, we observe that invasion often occurs at the contact point between colliding colonies, rather than at their original centres (Figs~\ref{fig:collision}B, C) - in good agreement with the experimental results of Fig.~\ref{fig:collision}A.

\section{Discussion}

Bacterial microcolonies provide attractive model systems for the study of multicellular assembly processes, because their development can be tracked in the lab at the level of individual cells, they are easy to manipulate (both physically and genetically) and, at least for {\em{E. coli}}, relatively few factors are at play. As well as mimicking eukaryotic  assembly processes such as
 the development of animal organs or tumour growth, bacterial communities provide an important test system for investigating the role of spatial structure in evolutionary processes such as the evolution of drug resistance, both in bacteria and in cancer~\cite{lambert_analogy_2011,greulich_mutational_2012}. Understanding microcolony growth also provides direct insights into the processes at play in the  early stages of bacterial biofilm formation and development.

Here, we have investigated the 2d to 3d transition that happens when an {\em{Escherichia coli}} microcolony invades a soft agarose surface. This transition provides a sensitive probe of the interactions both within the microcolony and with its environment, and also mimics  clinically and industrially relevant situations such as the invasion of soft tissue by bacteria growing on the surface of a medical implant or catheter,  and the invasion of foodstuff by pathogenic bacteria.

Our experimental results depend in a complex way on the elasticity of the agarose; the density of cells within the microcolony increases with the agarose concentration while the colony area at which the 2d to 3d transition happens peaks at an intermediate concentration. Using phenomenological theory, we show that the compression of the agarose by the microcolony and frictional forces between the bacteria and the agarose both play key roles. Combining this theory with individual-based computer simulations, we show that  mechanical interactions can explain all the observed features of colony growth, and of the 2d to 3d transition, observed in our experiments. The essential ingredients that we need in our simulations to reproduce the experimental data are: (1)~vertical and radial compression forces proportional to the elastic modulus of the agarose $E_{\rm a}$; (2)~a friction coefficient which decreases for low concentrations $C_{\rm a}$ down to an asymptotic value at high concentrations; (3)~a mechanism by which the cell packing can increase upon compression (in our simulations this can happen either by bending  or by longitudinal or transverse compression of the cells). We do not need to include any biochemical factors to reproduce our experimental results;  indeed these appear to be ruled out by our experimental observations for  colliding colonies.

Recently, individual-based computer simulations, in which rigid, rod-shaped bacteria interact via simple repulsive forces,  have been successful in reproducing several features of the morphology and dynamics of bacterial communities~\cite{volfson_biomechanical_2008,boyer_buckling_2011,haseloff2012,waclaw2013}. Our simulations go a step further by including the elasticity of the bacterial cells and the elastic interactions between the bacteria and the surrounding agarose medium. These features turn out to be essential in explaining the key features of the 2d to 3d transition, suggesting that elastic forces may also play an important role in other aspects of bacterial colony morphology. Our results also highlight the  key role of frictional forces, which are often treated only crudely in individual-based models, and have not, so far, been studied experimentally in any detail. Our analysis allows us to make a non-trivial prediction about the concentration-dependence of the friction coefficient between an {\em{E. coli}} bacterium and  agarose and glass surfaces, which should be testable using existing optical tweezers methods.

Our simulations have two unknown parameters: $k_0$, the value of the friction coefficient at agarose concentration $1.5\%$wt, and $E_{\rm b}$, the elastic modulus of a bacterial cell. The best match between our simulations and the experimental data is obtained for $E_{\rm b}=375$\,kPa. This is in fact in agreement with a value obtained recently in a different experiment by Tuson {\em{et al.}} \cite{agar_E}. These authors measured the elastic modulus of the  {\it E. coli} cell wall to be $E_{\rm wall} = 50 - 150$\,MPa. Assuming that the cell can be modelled by an empty cylinder of diameter is $d=1.4\,\mu$m and wall thickness $h=4\,\mu$m, one can compute the effective Young modulus of the bacterial cell in the longitudinal direction as $E_{\rm b} \approx 2h E_{\rm wall}/d = 285 - 860$\,kPa; our fitted value falls within this range.

The experimental study, supported by the simulation, convincingly points to the importance of mechanics in the invasion transition of the growing bacterium colony, as opposed to hypothetical biochemical effects acting on the bacteria or on the agarose structure. In future work we will explore in greater detail the transmission of force through the colony, which should be experimentally accessible by tracking tracer particles in the agar, as is routinely done in tracking forces generated by cellular tissues~\cite{Nelson12}.

The transition to bulk growth in {\em{E. coli}} microcolonies is a significant limitation in microscopic studies of single cell physiology and gene regulation, as well as limiting potential designs for bacteria-based bio-sensors. Our results show that this transition cannot easily be prevented by  changing the elasticity of the gel material. More broadly, this work suggests that invasion of a soft material, whether a foodstuff or an animal tissue, by a growing mass of cells, whether a bacterial colony or a cancer tumour, is likely to depend in a highly non-trivial way on the elastic properties of the material being invaded. Our work also highlights the urgent need for more single-cell measurements of the mechanical properties of bacteria, and the mechanical interactions between bacteria and their environment.

\section*{Appendix}
\renewcommand{\thesection}{Appendix \arabic{section}}
\setcounter{figure}{0}
\setcounter{equation}{0}
\setcounter{section}{0}
\renewcommand{\figurename}{Appendix Figure}
\renewcommand{\thefigure}{A\arabic{figure}}
\renewcommand{\thetable}{A\arabic{table}}
\renewcommand{\theequation}{A\arabic{equation}}

\subsection{Radial contribution due to the agarose elasticity.}
Here we show that the radial compression force acting on bacteria at the edge of the microcolony is approximately $E_{\rm a}d/(2\pi R)$. As in the main text, we model the microcolony as a rigid disk, pressed vertically into the agarose. The energy required in this process is, according to Eq.~(\ref{eq:vert}),
\bq
	\int_0^d dh \int_0^{R} dr \, 2\pi r \frac{E_{\rm a} h}{\pi\sqrt{R^2-r^2}} = E_{\rm a} R d^2. \label{eqs1}
\eq
Let us now imagine, that, instead of pressing the disk of constant radius into the agarose, we first press an infinitesimally thin cylinder to depth $d$ and then expand it from radius zero to radius $R$. The energy required for this process is
\bq
	\int_0^R dr \, 2\pi r d \, f_{\rm r,edge}(r), \label{eqs2}
\eq
where $f_{\rm r, edge}$ is the radial compression force per unit area, acting at the boundary and opposing the expansion. By equating (\ref{eqs1}) and (\ref{eqs2}) and differentiating with respect to $R$, we obtain that $f_{\rm r,edge}(R)=E_{\rm a}d/(2\pi R)$.

\begin{figure*}[t!]
	\includegraphics[width=14cm]{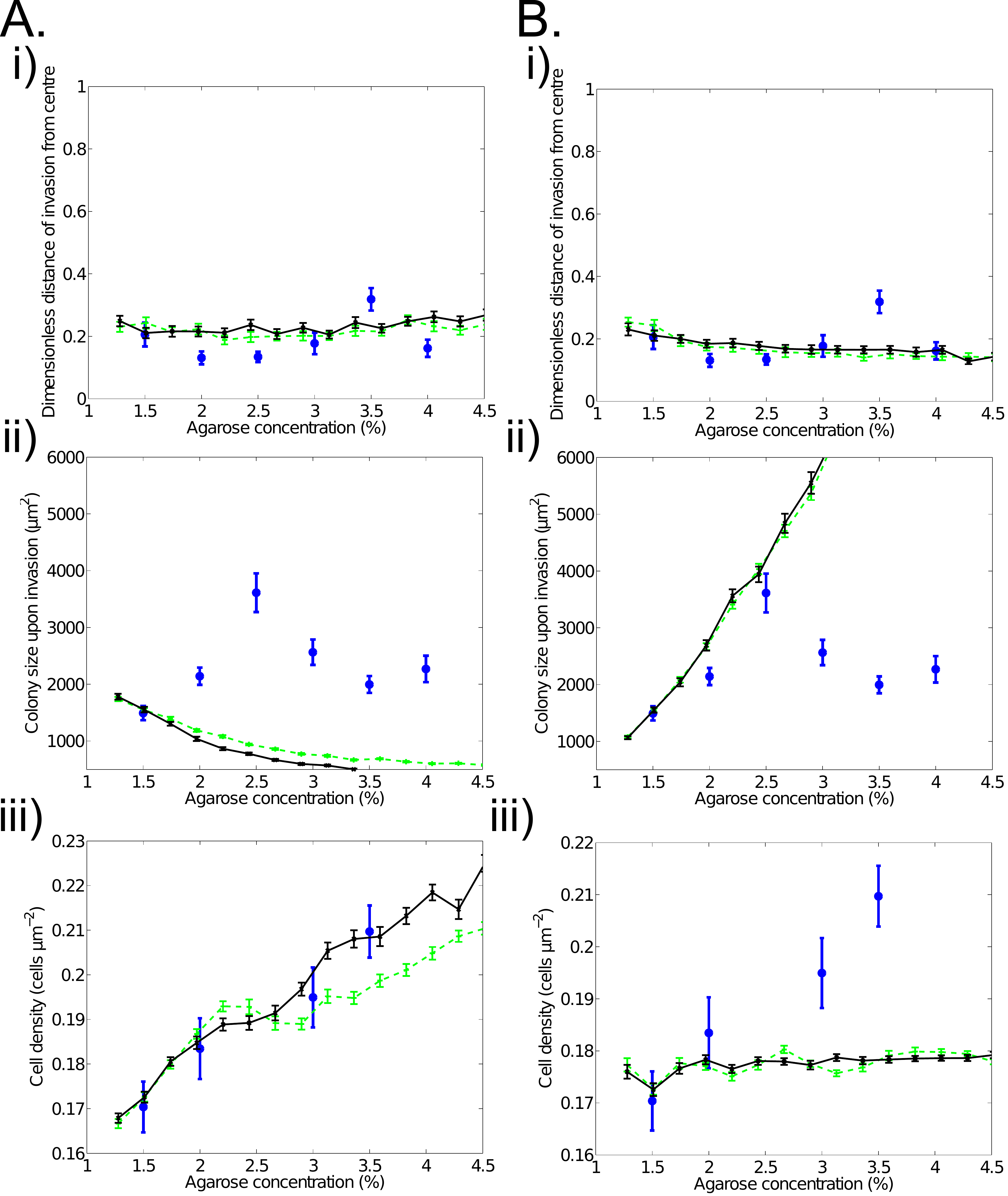}
	\caption{\label{fig:sf1}\textbf{Changing the dependence of the friction coefficient on the agarose concentration affects whether the simulations match the experimental data.} In all figures dark solid lines correspond to $\alpha=1$ and  dashed lines to $\alpha=0.4$ from Eq.~\ref{eq:hor}. Circles show experimental data. A: Simulations with \textit{constant friction coefficient} ($k=0.7$): (i)~The dimensionless buckling distance, which matches the experiments well. (ii)~The colony area upon invasion, which does not match the experimental data. (iii)~The cell density at the transition, which also matches well. B: Simulations with a \textit{friction coefficient that is inversely proportional to the agarose concentration} ($k=0.7\times 148/E_a$): (i)~The dimensionless buckling distance, which again matches well. (ii)~The colony area upon invasion, which matches well up to 2.5\%. (iii)~The cell density, which does not match the experimental data. To fully match the experimental data, see the form of Eq.~\ref{eq:kk} in the main text.}
\end{figure*}

\subsection{Constant friction coefficient $k$, and $k\sim 1/E_{\rm a}$}
In this section we show that if we assume either a constant friction coefficient $k$, or a friction coefficient that decreases inversely proportional to the agarose elastic modulus, $k\sim 1/E_{\rm a}$, then our simulations do not reproduce our experimental data. In Fig. \ref{fig:sf1}(A) we present the results for $k=\rm const = 0.7$, with the other simulation parameters chosen so that the microcolony area upon invasion agrees with the experimental result  for $1.5\%$ agarose. Although both the dimensionless distance of the buckling point from the centre and  the cell density at the transition agree well with the data, the microcolony area at the transition decays monotonically with increasing agarose concentration, failing to reproduce the peak seen in the experiments. The situation is slightly better for $k = 0.7 \times 148{\rm kPa}/E_{\rm a}$, Fig.~\ref{fig:sf1}(B), for which the area at the transition matches the experimental data up to the peak, but deviates from it for $>2.5\%$ agarose. These results clearly show that, in order to reproduce our experimental results, the friction coefficient can be neither constant nor monotonically decreasing with the agarose elastic modulus.

\begin{figure*}[t!]
	\includegraphics[width=16cm]{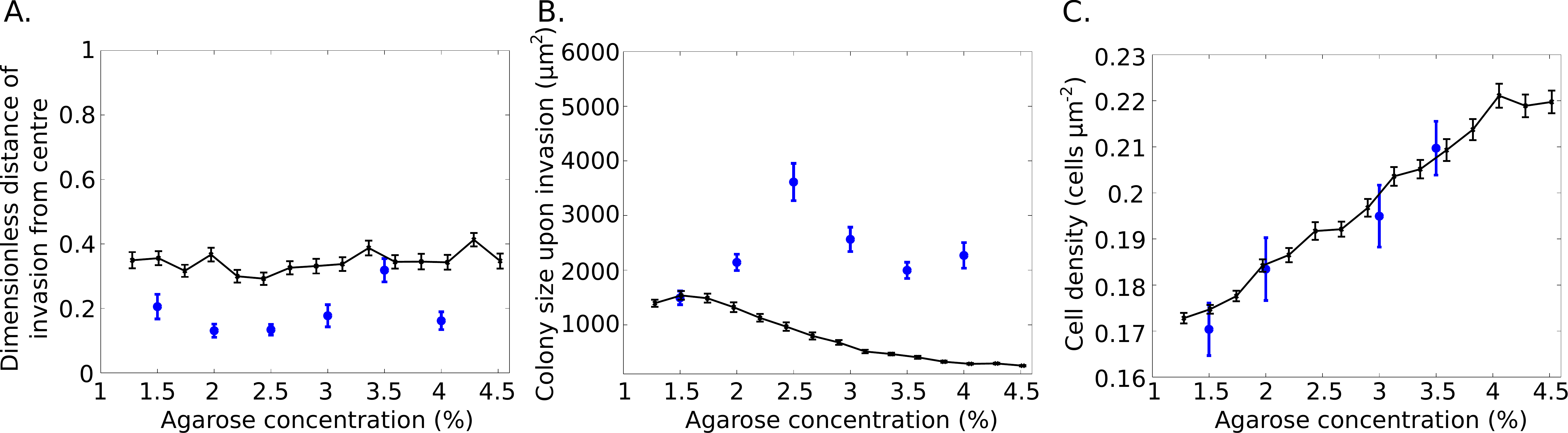}\\ 
\includegraphics[width=16cm]{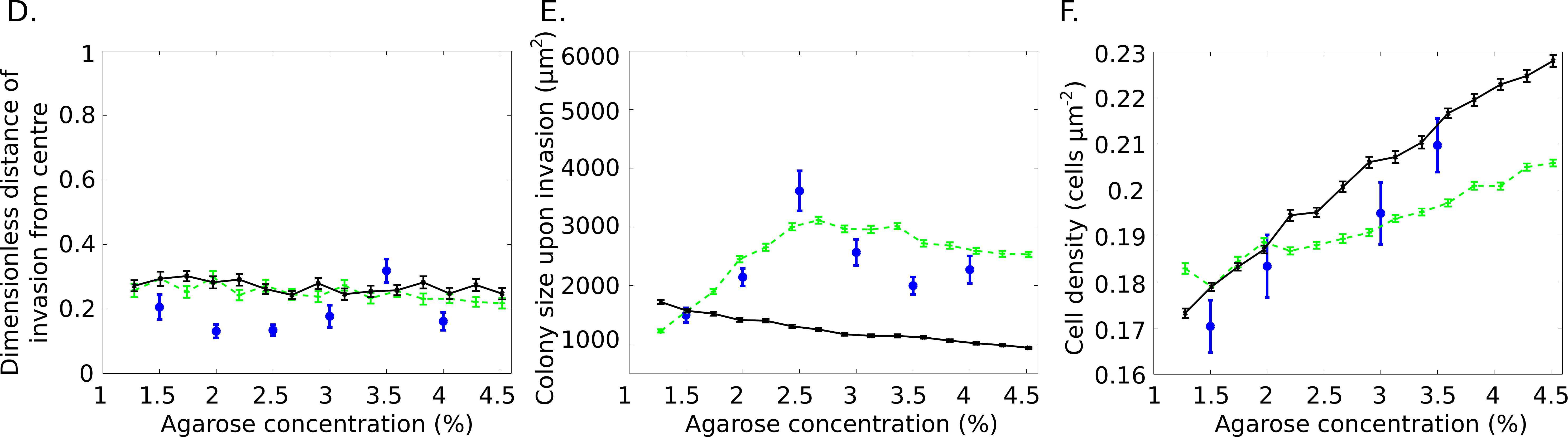}\\ 
	\caption{\label{fig:rad_only}\textbf{(A,B,C)~If friction is negligible, boundary-induced compression of the colony by the agarose cannot explain the experimental data}. The results of computer simulations (lines) where radial compression caused by the agarose acting on peripheral cells dominates friction from the substrate. Solid markers are experimental data. Apart from the density of the bacteria, which agrees with the experimental data for $E_{\rm b}=750$\,kPa, neither the area upon invasion nor the reduced distance can reproduce the corresponding data. This suggests that in our experiments,  radial compression due to the agarose is less important than radial  friction.
(D,E,F)~\textbf{``Stokesian'' friction proportional to velocity leads to similar results as velocity-independent friction}. Area (D), reduced distance (E), and cell density upon invasion (F), for constant $\kappa$ (solid line) and $\kappa$ depending on $C_{\rm a}$ as in Eq.~(\ref{eq:kk}) (dashed line). Solid  circles show experimental data.}
%
%
\end{figure*}

\subsection{Simulations of colliding colonies}
We simulate the growth of two colonies that are initially separated by a distance of 10\,$\mu$m or $30\,\mu$m using the same method as before, with the only exception being the calculation of the vertical and radial compression forces. Instead approximating the colony by a rigid disk with radius $R$, we use two disks of radii $R_1$ and $R_2$, centered at the initial positions of the two colonies and having same areas as the colonies. A bacterium is assumed to belong to disk 1 if its distance from the centre of disk 1 is smaller than from disk 2, and vice versa. Each bacterium experiences vertical compression given by Eq.~(\ref{eq:vert}), with $R$ replaced either by $R_1$ or $R_2$, depending on whether the bacterium is closer to the centre of disk 1 or disk 2. The radial compression force acting on the boundary of each sub-colony is calculated accordingly (i.e. using Eq.~(\ref{eq:rad}) but for two colonies).

\subsection{Radial compression larger than friction}
To test whether our data are compatible with a situation where the inward radial compression force from the agarose dominates over the radial frictional forces,  we performed simulations with a smaller friction coefficient $k=0.1$, and a larger radial compression (by a factor of 10 compared to our standard simulations). With this parameter set, the radial compression force  was almost 2 orders of magnitude  stronger than the friction force. Figure~\ref{fig:rad_only}(A-C) shows that in this case, the dimensionless distance of invasion becomes almost twice as large as it is in the experiments. This suggests that radial compression does not dominate in our system.

\subsection{Velocity-dependent friction}
We have also carried out simulations in which we assumed the friction coefficient to be proportional to the cells' velocity along the agar surface, $k=\kappa v$, with proportionality coefficient $\kappa$. In these simulations,  as before, we either assumed that $\kappa$ was constant, or that it depended non-monotonically on $C_{\rm a}$ in the same way as for the static friction coefficient $k$ in our previous simulations (Eq.~(\ref{eq:kk})).
Figure~\ref{fig:rad_only}(D,E,F) shows that all three quantities that we are interested in (area, distance, and density at transition) behave in a similar way as in our simulations with velocity-independent friction (compare to Figs.~\ref{fig:results} and \ref{fig:sf1}). The agreement with the experimental data is, however, a little worse than for the static friction.

\section*{Acknowledgments}
We thank J. Cholewa-Waclaw for many useful discussions regarding embryonic development and cell differentiation, M. Cosentino Lagomarsino and M. Osella for suggesting this problem, A. Javer for practical help, and D. P. Lloyd for sharing with us the results of his related experiments.
RJA was supported by a Royal Society University Research Fellowship.
BW was supported by the Leverhulme Trust Early Career Fellowship.  MAAG was funded by EPSRC.  PC acknowledges funding from the
International Human Frontier Science Program Organization, grant RGY0069/2009-C. RJA acknowledges funding from the same source, grant RGY0081/2012.




\end{document}